\documentclass[12pt,a4paper,twoside]{article}        
\usepackage[english]{babel}                          
\usepackage[cp1251]{inputenc}                        
\usepackage{mmgart-pdftex}                           

\begin{document}

\setcounter{page}{20}                                
\thispagestyle{empty}                                
\begin{heading}                                      
{Volume\;5,\, N{o}\;2,\, p.\,20 -- 29\, (2017)}      
{}                                                   
\end{heading}                                        

\begin{Title}
On the charge-to-mass ratio for self-gravitating systems of scalar and electromagnetic fields
\end{Title}

\begin{center}
\Author{a}{P. V. Kratovich}
\and
\Author{b}{Ju. V. Tchemarina}
\end{center}


\begin{flushleft}
\Address{}{Faculty of Mathematics, Tver State University, Sadovyi per. 35, Tver, Russia, 170002}


\Email{$^a$\,kratovich.pv@tversu.ru\,,$^b$\,chemarina.yv@tversu.ru}
\end{flushleft}

\Headers{P.\,V.\,Kratovich and Ju.\,V.\,Tchemarina}
{On the charge-to-mass ratio}

\begin{flushleft}                                 
\small\it Received 10 August 2017, in final form  
30 August.\ Published 31 August 2017.             
\end{flushleft}                                   


\Thanks{\mbox{}\\
\copyright\,The author(s) 2017. \ Published by Tver State University, Tver, Russia}
\renewcommand{\thefootnote}{\arabic{footnote}}
\setcounter{footnote}{0}

\Abstract{We prove that in the family of static, asymptotically flat, spherically symmetric scalar hairy black holes with the central electric charge, the set of the charge-to-mass ratios has the exact upper bound
$3\sqrt{2}/4\approx1.06$.}

\Keywords{scalar hairy black hole, charge-to-mass ratio}

\MSC{83C22, 83C57}


\newpage                               
\renewcommand{\baselinestretch}{1.1}   


\section{Introduction}

The investigation of compact self-gravitating configurations of physical fields has a long history~\cite{Fisher1948, Bergmann, Wyman1981, Bronnikov1973} and plays an important role in the modern theories of gravity and astrophysics. In particular, static self-gravitating real scalar fields are actively applied in the mathematical modelling of unusual hypothetical particle-like objects both in particle physics~\cite{Ellis, Rybakov} and in astrophysics and cosmology~\cite{Schunk, MatosGuzman2001, SchunckMielke2003}. From a physical point of view, a scalar field can be considered both as a fundamental natural field and a phenomenological construction when, e.g., dark matter is modelled.

In this article we deal with static, asymptotically flat, spherically symmetric self-gravitating configurations of a real scalar field minimally coupled to gravity and the electromagnetic field; we also assume that the scalar field is nonlinear with an arbitrary self-interaction potential. From a purely mathematical point of view, these configurations are solutions of the Einstein-Maxwell-Klein-Gordon equations. At the classical level, the no-hair theorem~\cite{MayoBekenstein} explicitly forbids their existence for black holes with self-interaction scalar field potentials being nonnegative everywhere outside of the event horizons. However, this restriction is not essential at the quantum level (quantum correction can be very large for particle-like solutions) and, also, is not true for naked singularities.

For the system under consideration, the field equations can be analytically integrated only for very special self-interaction potentials. On the other hand, even if scalar fields exist in nature, we do not know the explicit form of the potential. We are studying the scalar field configurations and their properties using the so-called 'inverse problem method for self-gravitating spherically symmetric scalar fields' developed in Refs.~\cite{BechmannLechtenfeld1995, DennhardtLechtenfeld1998, BronnikovShikin2002}. This method leads to the 'general solution' of the Einstein-Klein-Gordon equations in the form of quadrature formulas~\cite{Tchemarina2009, Azreg-Ainou2010, Solovyev2012}. It allows us, in some sense, to consider the configurations with all the admissible self-interaction potentials at the same time.

It is well-known that the family of vacuum charged black holes described by the Reissner-Nordstr\"{o}m solution has the exact upper bound $|q|/m\leqslant1$ for the charge-to-mass ratios~\cite{Chandrasekhar}, while for the corresponding family of charged naked singularities the inequality $|q|/m>1$ holds. It is obvious that a vacuum charged black holes cannot be considered as a model of any elementary particle since, for example, protons have the charge-to-mass ratio of the order of $10^{18}$. In this connection, there a natural question arises whether black holes with scalar hair have the ratio $q/m\gg1$. In this article we study the question of possible values of the charge-to-mass ratio for the systems under consideration. Our main goals are to show that these values are bounded from above and to find the corresponding exact upper bound for the black hole configurations. We give an exact formulation of the main result in the form of a theorem in Section~\ref{S2} and prove it in Section~\ref{S3}.

We adopt the spacetime metric signature $(+\,-\,-\,-)$, define the curvature tensor so that $R^i_{\,jkl}=\partial_k\Gamma^i_{\,jl}-\ldots\,$, and use geometric units in which $G = c = 1$.

\section{The main result}
\label{S2}

First, it is necessary to specify the class of spacetimes that we deal with as much as possible. The action of the gravitating system of a nonlinear real scalar field and the electromagnetic field, which assumed both to be minimally coupled to gravity, is
\begin{equation}\label{action}
\Sigma=\frac{1}{8\pi}\int\!\left(-\frac{1}{2}S\,+\, \left\langle d\phi,d\phi\right\rangle\,- 2\,V(\phi)\,-\frac{1}{2} \left\langle \mathcal{F},\mathcal{F}\right\rangle\right) \mathop{\sqrt[]{|g|}}d^{4}x\,,
\end{equation}
where $S$ is the scalar curvature, $\mathcal{F}$ is the electromagnetic field 2-form, $V(\phi)$  is the self-interaction potential of $\phi$, and the angle brackets denote the inner product induced by the spacetime metric.

We write the spherically symmetric spacetime metric in the form
\begin{equation}\label{metric}
ds^2=f(C)\:\!\mathrm{e}^{\,2F(C)}dt^2\,-\, \frac{dC^2}{f(C)}\,-\, C^{2}\!\left(d\theta^2\,+\,\sin^{2}\!\theta d\varphi^2\right).
\end{equation}
It is obvious that $f(C)=-\langle{}dC,dC\rangle$ and the solutions of the equation $f(C)=0$ define hypersurfaces on which the 1-form $dC$ becomes null. In particular, it is true on event horizons and hence the function $f(C)$ will be referred to as the characteristic function.
We use the symbol $C$ for the radial coordinate in order to have the future possibility of changing the coordinate condition, $C\rightarrow{}C(r),\; dC\rightarrow{}(dC/dr)dr$, with minimal changes in the basic formulae.

The Einstein-Klein-Gordon equations for the action~(\ref{action}) and the metric~(\ref{metric}) are written, for example, in Ref.~\cite{Solovyev2012}. We use the inverse problem method mentioned above in the introduction and the quadrature formulae obtained in Ref.~\cite{Solovyev2012} in the form (the prime denotes a derivative with respect to $C$)
\begin{equation}\label{F}
F\,=\,-\,\int\limits_{\!C}^{\;\infty} \left(\phi\,'\right)^{2}C\,dC\,,
\end{equation}
\begin{equation}\label{QP}
Q\,=\,C+\,\int\limits_{\!C}^{\;\infty} \left(1-\mathrm{e}^{F}\right)dC\,,\quad P\,=\,\int\limits_{\!C}^{\;\infty} \frac{\mathrm{e}^{F}}{C^{2}}\,dC\,,
\end{equation}
\begin{equation}\label{f}
f\,=\,2C^2\mathrm{e}^{-2F} \int\limits_{\!C}^{\;\infty} \frac{(Q\,-\,3\,m\,+\,2q^{2}P)\, \mathrm{e}^{F}}{C^{4}}\,dC\,,
\end{equation}
\begin{equation}\label{V}
\tilde{V}(C)\,=\,\frac{1}{2C^{2}}\! \left(1\,-\,f\,-\,C^{2}\left(\phi\,'\right)^{2}f\,-\, Cf\,'\,-\,\frac{q^{2}}{C^{2}}\right),
\end{equation}
\begin{equation}\label{qF}
\mathcal{F}\,=\, q\,\frac{\,\mathrm{e}^{F}}{C^2}\,dt\wedge dC\,,
\end{equation}
where $q$ and $m$ are arbitrary constants and the corresponding spacetime is asymptotically flat if and only if the scalar field has the asymptotic behaviour
\begin{equation}\label{asympt}
\phi\,'\,\equiv\,\frac{d\phi}{dC}\,=\, O\!\left(C^{-3/2\,-\,\alpha}\right),
\quad C\rightarrow +\infty \,,\quad \alpha>0\,.
\end{equation}
The method works as follows: choosing a monotonic field function $\phi(C)$, one finds successively the metric function $F(C)$, the functions $Q$ and $P$, and the metric function $f(C)$ by direct integration in~(\ref{F}),~(\ref{QP}), and~(\ref{f}) respectively; the self-interaction potential can be found as a function of $C$ from~(\ref{V}) and then, taking into account monotonicity of $\phi(C)$, as a function of $\phi$; the 2-form $\mathcal{F}$ is given by~(\ref{qF}). Note that the integral formulae~(\ref{F}) ---~(\ref{V}) are true in general, not only when $\phi(C)$ is a monotonic function. Therefore, in the considered family, each configuration of charge $q$ and mass $m$ depends also on the arbitrariness in the choice of the field function $\phi$ which obeys the asymptotic condition~(\ref{asympt}) and determines the self-interaction potential.

It is important that the formulae~(\ref{F}) ---~(\ref{f}) allow us to come to some conclusions about these configurations without any reference to the explicit form of the self-interaction potential. In particular, such a conclusion --- the main result of this article --- is given in the following theorem.
\begin{theorem}\label{th}
In the family of static, asymptotically flat, spherically symmetric black holes that are consistent with the action~(\ref{action}), the set of the charge-to-mass ratios has the exact upper bound
\begin{equation}\label{Theor}
\frac{|q|}{m}\,<\,\frac{3\sqrt{2}}{4}\, \approx\,1,060660172\,.
\end{equation}
\end{theorem}

\section{The proof of the main result}
\label{S3}

In the simplest case of the Reissner-Nordstr\"{o}m solution, the characteristic function has the form~\cite{Chandrasekhar}
$$
f\,=\,1\,-\,\frac{2\,m}{C}\,+\,\frac{q^{2}}{C^{2}}\,.
$$
In the region of the parameters $|q|<m$, the equation $f(C)=0$ defines black holes with the event horizon and the internal Cauchy horizon at the hypersurfaces
$$
C_{\pm}=\,m\,\pm\,\sqrt{m^{2}\,-\,q^{2}}\,,
$$
while if $|q|=m$, the two horizons coincide and one has the extreme black holes with $f(m)=0$ and $f'(m)=0$ at their horizons; the region $|q|>m$ corresponds to the naked singularities.

It turns out that the scalar field configurations in question are qualitatively similar to the Reissner-Nordstr\"{o}m ones. If $C=C^*$ is the horizon radius of an extremal scalar field black holes, then the characteristic function obeys the conditions
$$
f(C^{*})\,=\,0\,,\quad f'(C^{*})\,=\,0\,.
$$
It is easy to see directly from the formulae~(\ref{F}) ---~(\ref{f}) that in this case
\begin{equation}\label{m}
\qquad m\,=\,\frac{1}{3X(C^*)}\left\{P(C^{*}) \int\limits_{\!C^*}^{\;\infty} \frac{\,Q\,\mathrm{e}^{F}}{C^{\,4}}\,dC\,-\, Q(C^{*})\int\limits_{\!C^*}^{\;\infty} \frac{\,P\,\mathrm{e}^{F}}{C^{\,4}}\,dC\right\},
\end{equation}
\begin{equation}\label{q}
q^{2}\,=\,\frac{1}{2X(C^*)}\left\{\; \int\limits_{\!C^*}^{\;\infty} \frac{Q\,\mathrm{e}^{F}}{C^{4}}\,dC\,-\, Q(C^{*})\int\limits_{\!C^*}^{\;\infty} \frac{\,\mathrm{e}^{F}}{C^{4}}\,dC\right\},\qquad
\end{equation}
where
\begin{equation}\label{X}
X(C^*)\,=\,{P(C^{*})\int\limits_{\!C^*}^{\;\infty} \frac{\,\mathrm{e}^{F}}{C^{\,4}}\,dC\,-\, \int\limits_{\!C^*}^{\;\infty} \frac{\,P\,\mathrm{e}^{F}}{C^{4}}\,dC}\,.
\end{equation}
Having established a field function $\phi(C)$, the formulas~(\ref{m})~---~(\ref{X}) allows one to find the mass and the charge of the configuration for a given horizon radius.

It follows from the formulae~(\ref{F}) and the condition~(\ref{qF}) that the metric function $F(C)$ is nondecreasing, $F(C)\leqslant0$ for all $C>0$, and
\begin{equation}\label{exp1}
0\,<\,\mathrm{e}^{F(C)}\,\leqslant\,1\,, \quad C \in \left(0;\,+\infty\right)\,,
\end{equation}
\begin{equation}\label{exp2}
\mathrm{e}^{F(C)}\,=\,1\,+\, O\left(C^{\,-\,1\,-\,2\alpha}\right),\quad C\rightarrow +\infty\,.
\end{equation}
Next, $Q(C)$ is a monotonically increasing function because $Q'(C)=\mathrm{e}^{F(C)}>0$, and, as follows from~(\ref{exp1}) and~(\ref{exp2}),
\begin{equation}\label{Q>C}
Q(C)\geq{}C>0\,,\quad C\in\left(0;\,+\infty\right).
\end{equation}
In the same way, $P(C)$ is a monotonically decreasing function, and for all $C>0$
\begin{equation}\label{P}
P(C)\,=\,\int\limits_{\!C}^{\;\infty} \frac{\mathrm{e}^{F}}{C^{2}}\,dC\;\leqslant\; \int\limits_{\!C}^{\;\infty}\frac{1}{C^{2}}\,dC\,=\, \frac{1}{C}\,.
\end{equation}

We also have
\begin{equation}\label{asymptQP}
Q(C)\,=\,C\,+\,O\left(C^{\,-\,2\alpha}\right),\quad P(C)\,=\,\frac{1}{C}\,+\,O\left(C^{\,-\,2\,-\, 2\alpha}\right),\quad C\rightarrow +\infty\,.
\end{equation}

In what follows we assume, for definiteness, that $q>0$.

Let $C^*$ be the event horizon radius, and $m$ the mass of a black hole, so that $f(C^*)=0$ and, as a consequence,
\begin{equation}\label{integrand}
\int\limits_{\!C^*}^{\;\infty}\frac{Q\,-\,3\,m\,-\, 2\,q^{2}\,P}{C^{\,4}}\,\mathrm{e}^{\,F}dC\,=\,0\,. \nonumber
\end{equation}
On the other hand, in the region $C\rightarrow +\infty$ we have
$$
Q(C)\,-\,3m\,+\,2q^{2}P(C)\,>\,0\,,
$$
therefore there exists a radius $\,C^{**}>C^*$ such that
\begin{equation}\label{C**}
Q(C^{**})\,-\,3m\,+\,2q^{2}P(C^{**})\,=\,0\,.\nonumber
\end{equation}
From this equality we find the mass-to-charge ratio in the form
\begin{equation}\label{mq}
\frac{m}{q}\,=\, \frac{1}{3} \left(\frac{Q(C^{**})}{q}\,+\,2qP(C^{**})\right)\,.
\end{equation}
Now we consider $C^{**}$ to be fixed and denote the right hand side in~(\ref{mq}) by $L(q)$. The function $L(q)$ has a unique minimum at
$$
q_{min}\,=\,\sqrt{\frac{Q(C^{**})}{2P(C^{**})}}\,,
$$
so that for all $q>0$
$$
L(q)\,\geqslant\,\,L_{min}\,=\,L(q_{min})\,=\, \frac{2}{3}\sqrt{2\,Q(C^{**})P(C^{**})}\,.
$$
Substituting the right-hand side of this inequality instead of $L(q)$ in the formula~(\ref{mq}), we obtain the estimate
\begin{equation}\label{m/q}
\frac{m}{q}\,\geqslant\,\frac{2\sqrt{2}}{3} \sqrt{Q(C^{**})P(C^{**})}\,.
\end{equation}
Next, it is easy to obtain directly from the formula~(\ref{QP}) the relation
$$
(QP)'(C)\,=\,\mathrm{e}^{\,F(C)}\left(P(C)\,-\, \frac{Q(C)}{C^{2}}\right).
$$
As a consequence of the inequalities~(\ref{Q>C}) and~(\ref{P}) we have
$$
P(C)\,-\,\frac{Q(C)}{C^{2}}\,\leqslant\,0\,,
$$
and $\,(QP)'(C)\,\leqslant\,0$; the latter means that $Q(C)P(C)\,$ is a nonincreasing function. Since, from~(\ref{asymptQP}),
\begin{equation}\label{}
\lim_{C\rightarrow\,+\infty}{(Q(C)P(C))}\,=\,1\,, \nonumber
\end{equation}
the inequality $Q(C)P(C)\,\geqslant\,1$ holds for all $C>0$ and, as follows from~(\ref{m/q}),
\begin{equation}\label{m-q}
\frac{m}{q}\,\geqslant\,\frac{2\sqrt{2}}{3}\,.
\end{equation}

Now we prove that the equality in~(\ref{m-q}) is impossible. Suppose,
on the contrary, that $m/q=2\sqrt{2}/3$. Then, as follows from~(\ref{m/q}), the inequality
$Q(C^{**})P(C^{**})\leqslant1\,$ holds. On the other hand, the condition $Q(C)P(C)\geqslant1$ implies
\begin{equation}\label{1}
Q(C)P(C)\,\equiv\,1 \quad \text{for all}\quad C \in \left[C^{\,**};\,+\infty\right)\,.
\end{equation}
Differentiating~(\ref{1}), we find
\begin{equation}\label{2}
P(C)\,-\,\frac{Q(C)}{C^{2}}\,\leqslant\,0 \quad \text{for all}\quad C \in \left[C^{\,**};\,+\infty\right)\,.
\end{equation}
Equations~(\ref{1}) and~(\ref{2}) yield
\begin{equation}\label{P=Q=}
P(C)\,=\,-\,\frac{1}{C}\,,\quad Q(C)\,=\,C \quad \mbox{for all}\quad C \in \left[C^{\,**};\,+\infty\right), \nonumber
\end{equation}
so that the relation~(\ref{m/q}) can be rewritten in the form
$$
\frac{2\sqrt{2}}{3}\,=\,\frac{1}{3} \left(\frac{C^{**}}{q}\,+\,\frac{2q}{C^{**}}\right).
$$
The latter, in turn, gives
\begin{equation}\label{}
q\,=\,\frac{\sqrt{2}\,C^{\,**}}{2}\,,\quad m\,=\,\frac{2\,C^{\,**}}{3}\,.
\end{equation}
In the region $C\in\left(0;\,+\infty\right)$, for the function $M(C)=Q(C)-3m-2q^{2}P(C)\,$ contained in the integrand in~(\ref{integrand}), we have the relations
$$
M(C)\,=\,Q(\,C)\,-\,2C^{**}\,-\, \left(C^{**}\right)^{2}P(C)\,,\quad
M'(C)\,=\,\mathrm{e}^{F(C)}\!\left(1\,-\, \frac{\left(C^{**}\right)^{2}}{C^{2}}\right),
$$
which show that $M(C)$ has the unique minimum point, $C=C^{**}$, in the region $C\in\left(0;\,+\infty\right)$, and, taking into account the relations~(\ref{P=Q=}), $M(C^{**})=0$. It implies successively that $M(C)>0$ for all $C\neq{}C^{**}$, the integrand in~(\ref{integrand}) is positive almost everywhere in the region $C\in\left(0;\,+\infty\right)$, and therefore the equality~(\ref{integrand}) cannot be fulfilled. We obtain a contradiction that proves the impossibility of the equality in~(\ref{m-q}) and the validity of the estimate~(\ref{Theor}).

Finally, we give an outline of the proof that the number $3\sqrt{2}/4$ in the estimate~(\ref{Theor}) is the exact upper bound for the possible values of $q/m$. A convenient and simple way to do this is to use a family of general solutions that belong to the space of distributions on $C\in\left(0;\,+\infty\right)$. Namely, we define the one-parameter family of the metric functions
\begin{equation}\label{family}
\mathrm{e}^{F(C)}=\begin{cases}
\dfrac{1}{a^{\,2}}\,,\;\;\mbox{if}\;\; C\,\in\,\left[0;\,a\right), \vphantom{\int\limits_a}\\
\;\;1\,,\,\;\;\text{if}\;\; C\,\in\,\left[a;\,+\infty\right),\quad 1<a<\infty\,,
\end{cases}
\end{equation}
and choose the mass $m$ and the charge $q$ so that the corresponding spacetimes will be extremal black holes with the horizon radius $C^{*}=1$. For a given value of $a$, the scalar field is a step function having some constant values, say $\phi_1$ and $\phi_2$, in the segments $\left[0;\,a\right)$ and $\left[a;\,+\infty\right)$ respectively. This function can be thought of as the limit, as $t\rightarrow0$, of the family
\begin{equation}\label{}
\phi(C)\,=\, \begin{cases}
\phi_1,\vphantom{\int\limits_a}\;\;C\in[0;\,a-t]\,;\\ \phi_2\dfrac{C-a+t}{2t}- \vphantom{\int\limits_a} \phi_1\dfrac{C-a-t}{2t}),\;\; C\in(a-t;\,a+t)\,;\\
\phi_2,\; \; C\in[a+t;\,\infty]\,. \nonumber
\end{cases}
\end{equation}

For the configurations defined by~(\ref{family}), the formulas~(\ref{QP}) yield
\begin{equation}\label{QC}
Q(C)\,=\begin{cases}
\dfrac{a^{3}-a+C}{a^{2}}\,, \quad C\,\in\,\left[0;\,a\right); \vphantom{\int\limits_a}\\
C\,,\quad C\,\in\,\left[a;\,+\infty\right),
\end{cases}
\end{equation}
\begin{equation}\label{PC}
P(C)\,=\begin{cases}
\dfrac{a^{2}C+a-C}{a^{3}C}\,,\quad C\,\in\,\left[0;\,a\right);\vphantom{\int\limits_a}\\
-\,\dfrac{1}{C}\,,\quad C\,\in\,\left[a;\,+\infty\right).
\end{cases}
\end{equation}
Substituting~(\ref{family}), (\ref{QC}), and~(\ref{PC}) in the formulas~(\ref{m}) and~(\ref{q}), we obtain
$$
\frac{q}{m}\,=\,\dfrac{3\,a^{5/2}\, \sqrt{\left(a^{5}\,+\,3\,a^{3}-\,2\,a^{2}\,-\, 3\,a\,+\,2\right)\left(2\,a^{4}+\,4\,a^{3}-\, 4\,a^{2}-\, 4\,a\,+\,3\right)}}{4\,a^{7}+6\,a^{6}-\,2\,a^{5}-\, 4\,a^{4}-\,5\,a^{3}+\,2\,a^{2}+\,3\,a-\,1}\,.
$$
In the limit $a\rightarrow\infty$, the latter relation gives
$$
\lim_{a\rightarrow \infty\,}{\frac{q}{m}}\,=\,\frac{3\,\sqrt{2}}{4}\,.
$$
In other words, for any $\varepsilon>0$ there exists a value of the parameter $a$ such that
$$
\frac{q}{m}\,>\,\frac{3\,\sqrt{2}}{4}\,-\,\varepsilon\,.
$$
Theorem~\ref{th} is proved.

\section{Conclusions}

In the vacuum case, the static charged black holes have the exact upper bound of the charge-to-mass ratios equal to one. If the mass $m$ is fixed and the charge is increasing to its maximum value $q=m$, then the event horizon radius is decreasing to its minimum value $C=m$ for the corresponding extreme black hole. Also, an uncharge, purely scalar field black hole of the mass $m$ has the event horizon radius which is strictly less than $2m$ and can be arbitrarily small~\cite{Solovyev2012}. In other words, both the scalar field and the electromagnetic field are purely repulsive and the balance is attained due to minimal coupling between the fields and gravity. In the light of these properties, the estimate of Theorem~\ref{th} seems to be surprising: one might expect that the upper bound for the charge-to-mass ratios would be much larger than one. In fact, however, it is approximately 1.06, which implies the corresponding consequences for some theories in particle physics and astrophysics.

\end{document}